\documentclass[english,aps,prstper,reprint,showpacs,titlepage,longbibliography]{revtex4-2}   
\usepackage[T1]{fontenc}	
\usepackage{geometry}		
\usepackage{graphicx}
\usepackage[above,below]{placeins}	
\usepackage{times}

\usepackage{hyperref}  
\hypersetup{colorlinks=true,urlcolor=blue,citecolor=blue,linkcolor=blue}   
\usepackage{enumitem}          
\setlist{nosep}                 

\begin{document}

\begin{titlepage}

\title{Who Gets to Do Physics? Occupational Stereotypes in AI-Generated Problem Sets}

 \author{Bilas Paul}
\email[Please address correspondence to ]{palb@farmingdale.edu}
\affiliation{SUNY Farmingdale State College, Farmingdale, NY, 11735, USA}

\begin{abstract}
\label{abstract}
\noindent 
As AI-generated problem sets gain traction in introductory physics courses, their technical correctness is well established — but the social assumptions embedded in their framing have gone largely unexamined.
This study analyzes 600 introductory physics problems generated by four AI systems — Grok~4, GPT-5.2, Claude Sonnet 4.6, and Gemini 3 Flash — across structured prompts involving occupations (CEO, Physicist, High School Teacher, Nurse, Construction Worker, and Migrant Worker).  Problems were coded  on five dimensions: hazard presence, hazard type, agency role,  cognitive role, and object ownership. While the physics content is technically sound across all platforms, our analysis reveals systematic occupational stratification in narrative framing. Hazardous scenarios were concentrated in Migrant Worker and Construction
Worker problems, with exposure-related hazards (electrocution, burns, radiation, heat or chemical exposure) especially concentrated in Migrant Worker problems. Passive-accident framing --- the persona as the
recipient of an injury --- appeared in one in eight Migrant Worker problems and never appeared for the Physicist, Teacher, or CEO. Possessive ownership language was reserved almost exclusively for the CEO. These patterns suggest that AI-generated physics problems can introduce surface-level diversity while reproducing occupational hierarchies in who acts, who owns, and who is placed at risk.  We discuss implications for physics teaching and offer simple screening strategies for instructors using AI-generated problems.
\end{abstract}

\maketitle
\end{titlepage}

\section{Introduction}
\label{intro}
Physics problems are never just about physics. Every time a student opens a problem set, they encounter not only forces, voltages, and velocities, but also a cast of characters --- people driving cars,
lifting objects, running experiments, or operating machinery. These contextual details are easy to overlook when grading for mathematical correctness, but research in physics education has shown
that they matter. Students are more likely to persist in physics when they can see themselves as the kinds of people who \textit{do} physics --- when physics is something done \textit{by} people like
them, not something that \textit{happens to} people like them~\cite{Hazari2000, Carlone2007}.

Artificial intelligence tools have made it easier than ever to generate large, varied problem sets in minutes. Many of us now routinely use tools like ChatGPT, Claude, or Gemini to produce
practice problems featuring diverse characters and workplace settings --- a welcome departure from the narrow range of contexts found in traditional textbooks~\cite{Gregorcic_2023, KASNECI2023102274,
MAROY2025UTI, Yeadon_2023}. As physics teachers, we appreciate the time these tools save and the relatability they can bring to our classrooms. On the surface, this looks like progress
toward more inclusive physics instruction. But does the diversity go deeper than the name of the character?

To find out, we generated 600 introductory physics problems across four major AI platforms using six occupational personas --- CEO, Physicist, High School Teacher, Nurse, Construction Worker, and
Migrant Worker --- and systematically examined how the narrative framing of each problem changed depending on who the character was. We were not evaluating the physics, which was generally correct throughout. Instead, we examined how the persona was positioned within the problem narrative: whether the character actively engaged with physical phenomena, observed them, or experienced them passively through accidents, hazards, or workplace risk.

Large language models are trained on vast corpora of human-generated text that reflect existing social hierarchies and occupational stereotypes~\cite{Bender2021, Blodgett2020}. When those models generate physics problems at scale, beyond supplying numerical exercises, they may reproduce narratives about who performs physics and who merely experiences it, automatically and without the editorial
oversight that governs traditional instructional materials. Our goal is to identify patterns that are easy to miss when reviewing problems for mathematical correctness alone, and to provide instructors with
practical strategies for evaluating AI-generated problems before they reach students.

\section{Materials and Methods}
\label{sec:methods}
In January 2026, we prompted four AI platforms --- Grok~4, GPT-5.2, Claude Sonnet~4.6, and Gemini~3 Flash --- with identical instructions to generate introductory physics problems for six occupational personas: CEO, Physicist, High School Teacher, Nurse, Construction Worker, and Migrant Worker. These personas were selected to represent a range of occupational roles associated with differing forms of social status, labor, and expertise. Each platform was asked to produce 25 problems per persona covering topics from algebra-based Physics I and II, including mechanics, fluids, waves, thermodynamics, electricity, magnetism, and optics. All platforms received the same prompt, with only the persona name changed across prompts. No constraints were placed on how the character should be portrayed, and the systems were not instructed to avoid hazards or specific narrative structures. The resulting corpus contained 100 problems per persona and 600 problems in total. All outputs were recorded without editing prior to analysis.

We developed a five-variable coding rubric to capture the narrative framing of each problem independently of its mathematical content. Each problem was coded once per variable based solely on how the scenario was described, not on the physics topic involved. The five variables were selected because they capture dimensions of narrative framing that physics education research has identified as consequential for student identity and belonging, even when such features go unnoticed during routine problem review~\cite{Hazari2000, Carlone2007, Rainey2018}.  Hazard presence and hazard type capture the level and nature of risk embedded in a scenario; agency role reflects whether the character initiates the physical action or experiences it; cognitive role distinguishes between analytical thinking and physical or passive involvement; and object relation captures whether the character is portrayed as owning or merely using equipment. Together, these variables provide a practical framework that instructors can apply when reviewing AI-generated problems for classroom use.

\section{Results}
\label{sec:results}
\subsection{Hazard Framing}
\label{sec:hazard_framing}
\begin{figure}[htb!]
\centering
\includegraphics[width=0.495\textwidth]{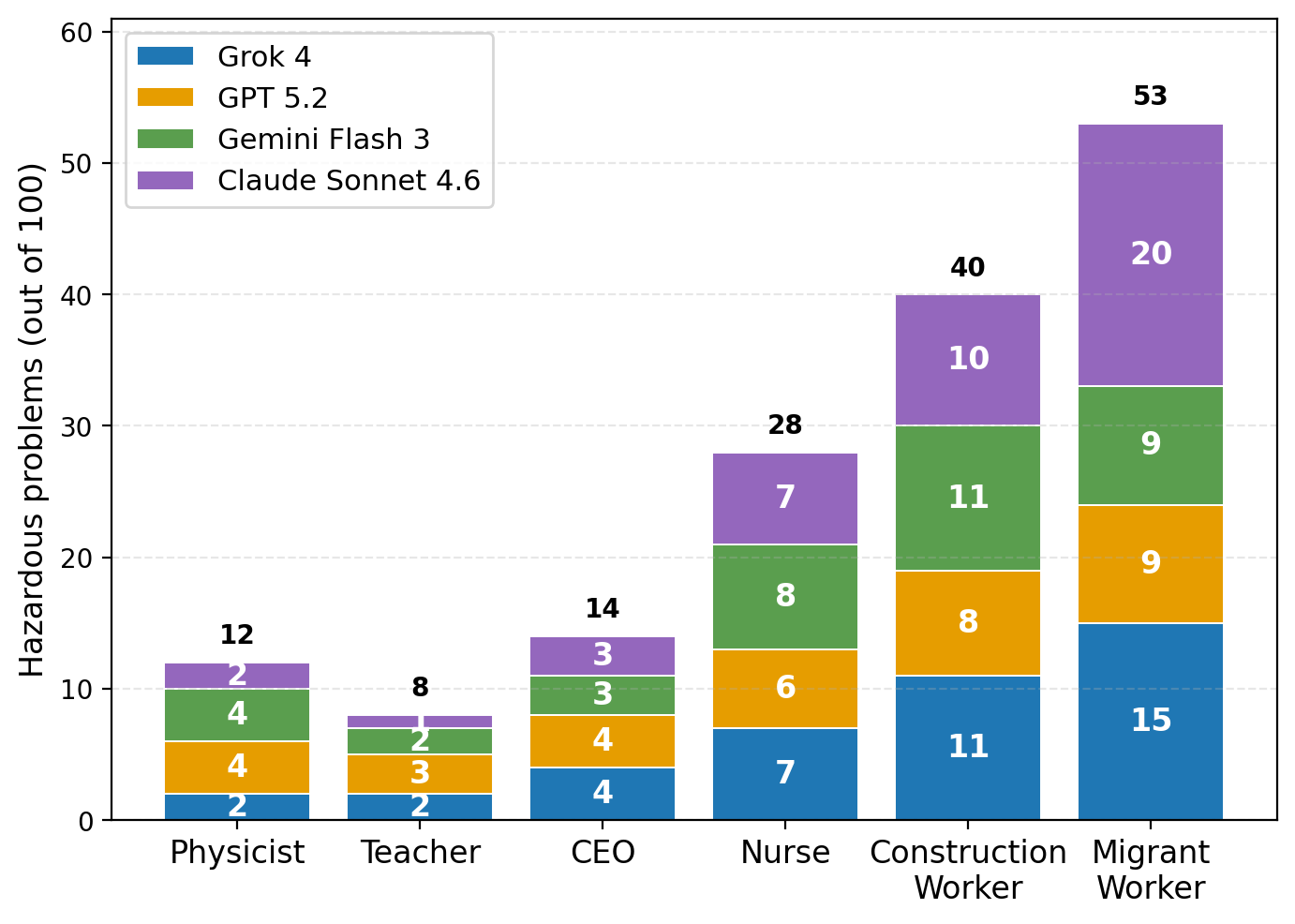}
\caption{Hazard exposure counts by occupational persona across four AI platforms (100 problems per persona, 25 per platform). Bars show problems coded with non-zero hazard presence, stacked by platform of origin. Professional personas consistently received the fewest hazardous problems, while the Migrant Worker accounted for the largest share across all platforms.}
\label{fig:hazard_framing}
\end{figure}

Hazard presence was scored on a five-level scale ranging from no hazard to severe physical risk.  Problems received a non-zero hazard code only when the scenario explicitly placed the persona in a potentially harmful physical situation.  Figure~\ref{fig:hazard_framing} shows the distribution of hazardous problems (\textit{hazard presence} $\geq 1$) by occupational persona, stacked by AI platform.  The pattern was systematic across platforms. The High School Teacher received the fewest hazardous problems (8 out of 100), followed by the Physicist (12) and the CEO (14). The Nurse occupied an intermediate position with 28 problems, reflecting clinical contexts involving patient care, medical procedures, and routine exposure to diagnostic devices. The Construction Worker reached 40, and the Migrant Worker reached 53 --- more than six times the hazard count of the High School Teacher and more than four times that of the Physicist.

The association between persona and hazard exposure was highly significant ($\chi^{2}=83.13$, $df=5$, $p<10^{-15}$). A focused comparison between the Migrant Worker and all other personas combined also remained significant ($\chi^{2}=44.54$, $p<10^{-10}$). The occupational ordering was preserved on every platform we tested, with the Migrant Worker receiving more hazardous problems than every professional persona regardless of which AI system generated them. Because many introductory physics topics naturally involve motion, forces, or electrical systems, hazardous scenarios could plausibly have appeared across multiple personas. Instead, they were concentrated disproportionately in problems featuring the Migrant Worker and, to a lesser extent, the Construction Worker.

\subsection{What Kind of Hazard?}
\label{sec:hazard_type}
\begin{figure}[htb!]
\centering
\includegraphics[width=0.495\textwidth]{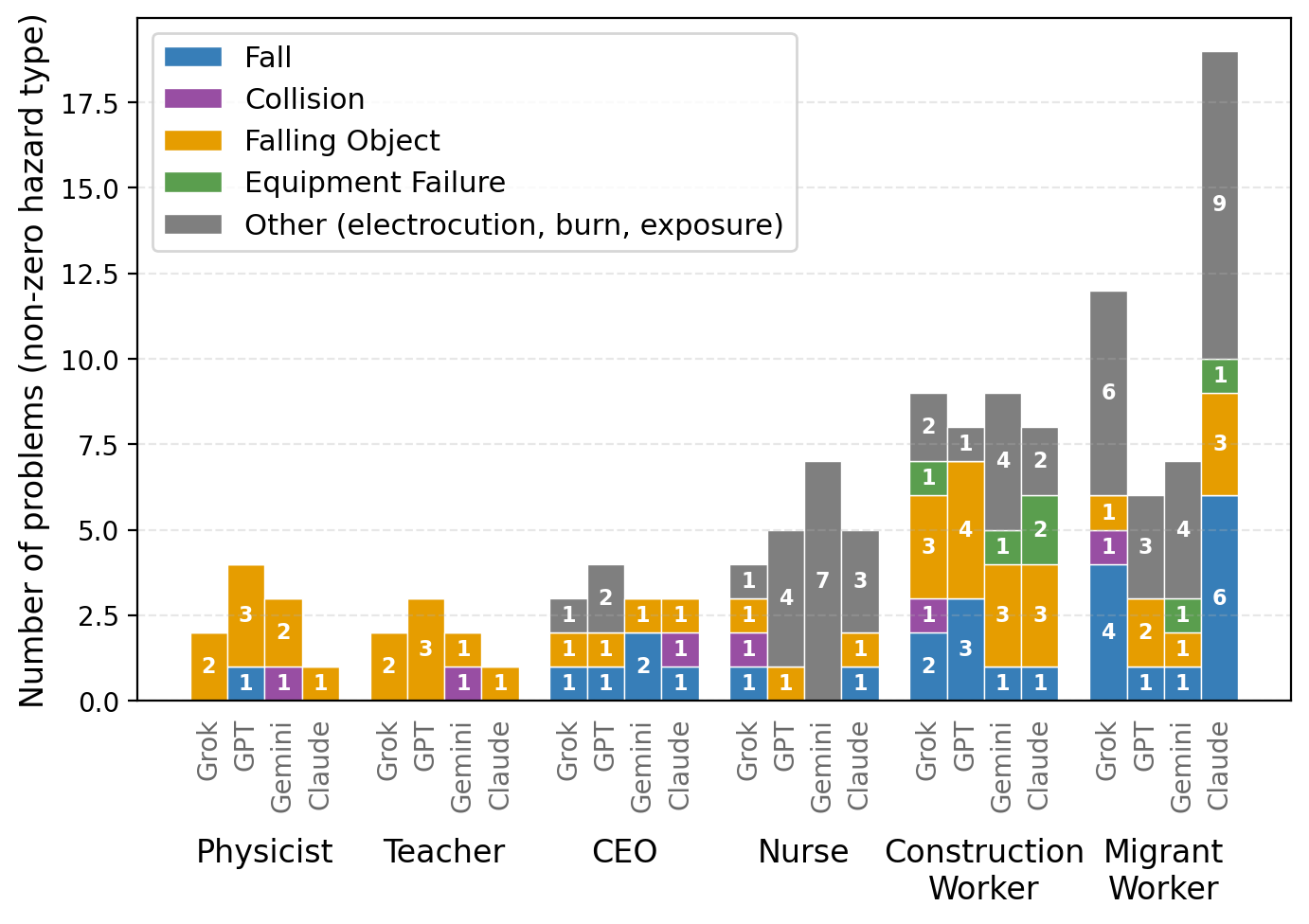}
\caption{Hazard type by persona and platform, restricted to problems with non-zero hazard type. Hazard types: 1 = fall, 2 = collision, 3 = falling object, 4 = equipment failure, 5 = other (electrocution, burn, radiation, chemical or heat exposure). Numbers inside segments indicate problem counts.}
\label{fig:hazard_type}
\end{figure}

Not all hazards are alike. Hazard type was scored on a six-level scale: 0 = none, 1 = fall, 2 = collision, 3 = falling object, 4 = equipment failure, 5 = other (electrocution, burn, radiation, chemical exposure, or heat exposure). Figure~\ref{fig:hazard_type} shows the breakdown by persona and platform, restricted to problems with a non-zero hazard type.

Hazard types were not distributed evenly across personas. Falling-object scenarios appeared across nearly all occupations, including the Physicist and High School Teacher, consistent with common introductory mechanics contexts. Collision scenarios were rare and roughly uniform across personas. The strongest differences appeared for equipment failure,  falls and exposure-related hazards. Equipment failure appeared only in personas whose work brings them into contact with operational equipment --- Construction Worker, and Migrant Worker --- and never in Physicist, Teacher, Nurse or CEO  scenarios. Fall scenarios appeared in 12 Migrant Worker, 7 Construction Worker, 5 CEO, 2 Nurse, 1 Physicist, and 0 High School Teacher problems --- but the CEO falls were qualitatively different from the rest: three of the five placed the CEO in recreational settings (sliding down a frictionless hill at a company retreat, for example), rather than depicting an occupational fall hazard. The Migrant Worker and Construction Worker falls, by contrast, almost always occurred at work. The sharpest stratification was in the ``other'' category. Of the 49 exposure-related problems, 22 involved the Migrant Worker, 15 the Nurse, 9 the Construction Worker, 3 the CEO, and zero the Physicist or Teacher. The Migrant Worker alone accounted for 45\% of all exposure-related hazards despite representing only one-sixth of the dataset (Fisher's exact test: odds ratio = 4.94, $p < 10^{-5}$). The Nurse and Migrant/Construction Worker exposure scenarios also differed qualitatively.  Nurse exposure scenarios typically involved professional use of medical equipment, such as defibrillators, MRI machines, or diagnostic devices. Migrant and Construction Worker exposure scenarios more often placed the persona as the object of risk within the work environment --- the same hazard code, but reflecting a vulnerable body rather than an expert operator.This distinction shows why hazard type, not just hazard frequency, matters for interpreting the generated problems.

\subsection{Who Acts and Who Is Acted Upon?}
\label{sec:agency}
\begin{figure}[htb!]
\centering
\includegraphics[width=0.495\textwidth]{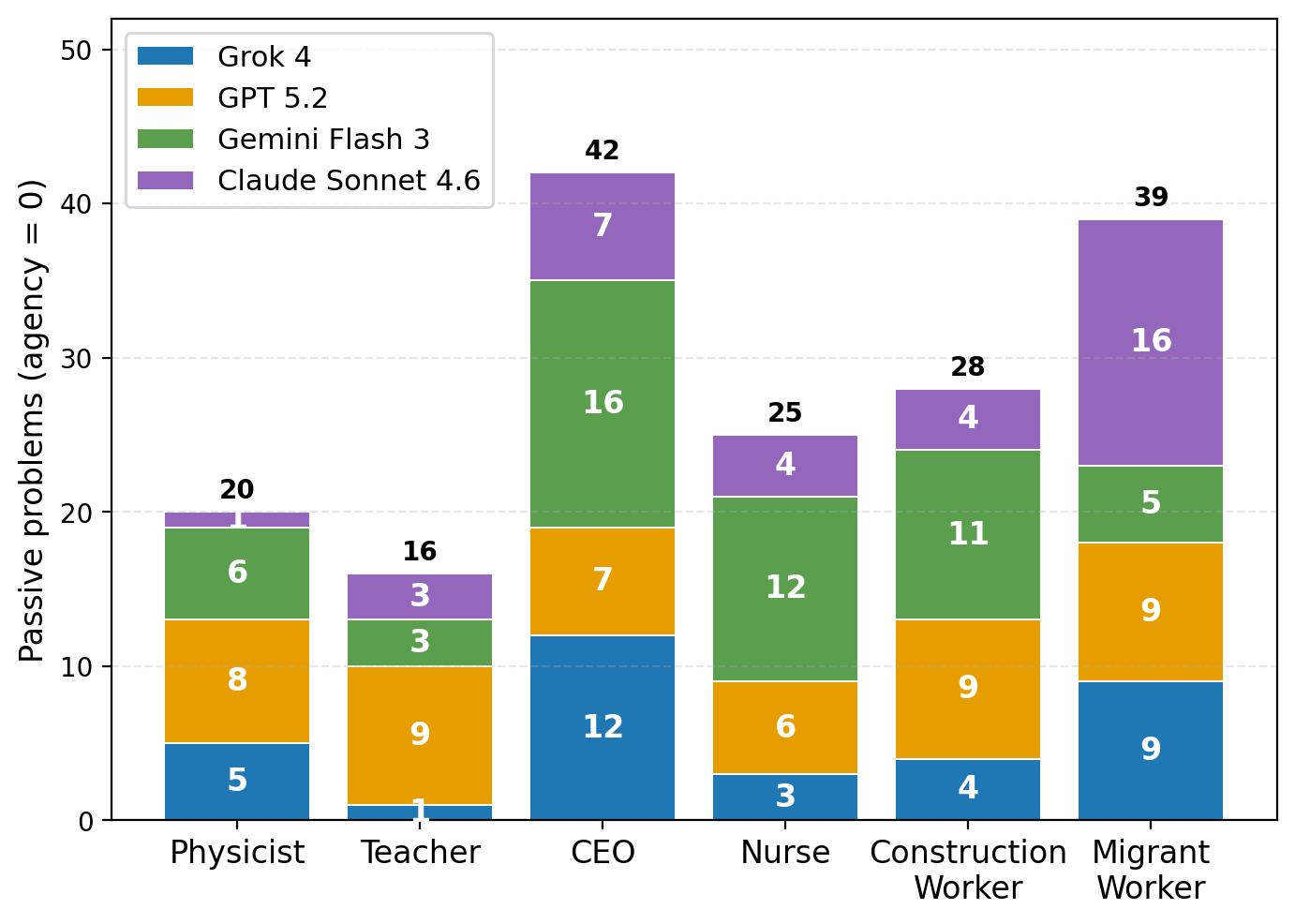}
\caption{Passive framings by occupational persona across four AI platforms. Bars are stacked by platform of origin. Counts reflect problems in which the persona did not initiate the physical action.}
\label{fig:agency}
\end{figure}

Agency captured whether the persona actively initiated the physical action or passively experienced it. Figure~\ref{fig:agency} shows the distribution of passive framings  by persona and platform.

Passive framings appeared across all personas but were unevenly distributed. The CEO received the most passive framings (42 out of 100), followed by the Migrant Worker (39), Construction Worker (28), Nurse (25), Physicist (20), and High School Teacher (16).  The CEO and Migrant Worker counts, however, reflected different narrative contexts. CEO passive framings typically involved observation or passive transport within safe environments, such as riding in an elevator, traveling on a boat, or hearing a siren from a stationary position. Migrant Worker passive framings more often involved exposure to workplace hazards, including falling from scaffolds, standing near dangerous equipment, or being exposed to heat and electrical systems. Thus, similar agency codes often corresponded to qualitatively different forms of narrative framing.

\subsection{Cognitive Role: Thinking vs.\ Experiencing}
\label{sec:cognitive}
\begin{figure}[htb!]
\centering
\includegraphics[width=0.495\textwidth]{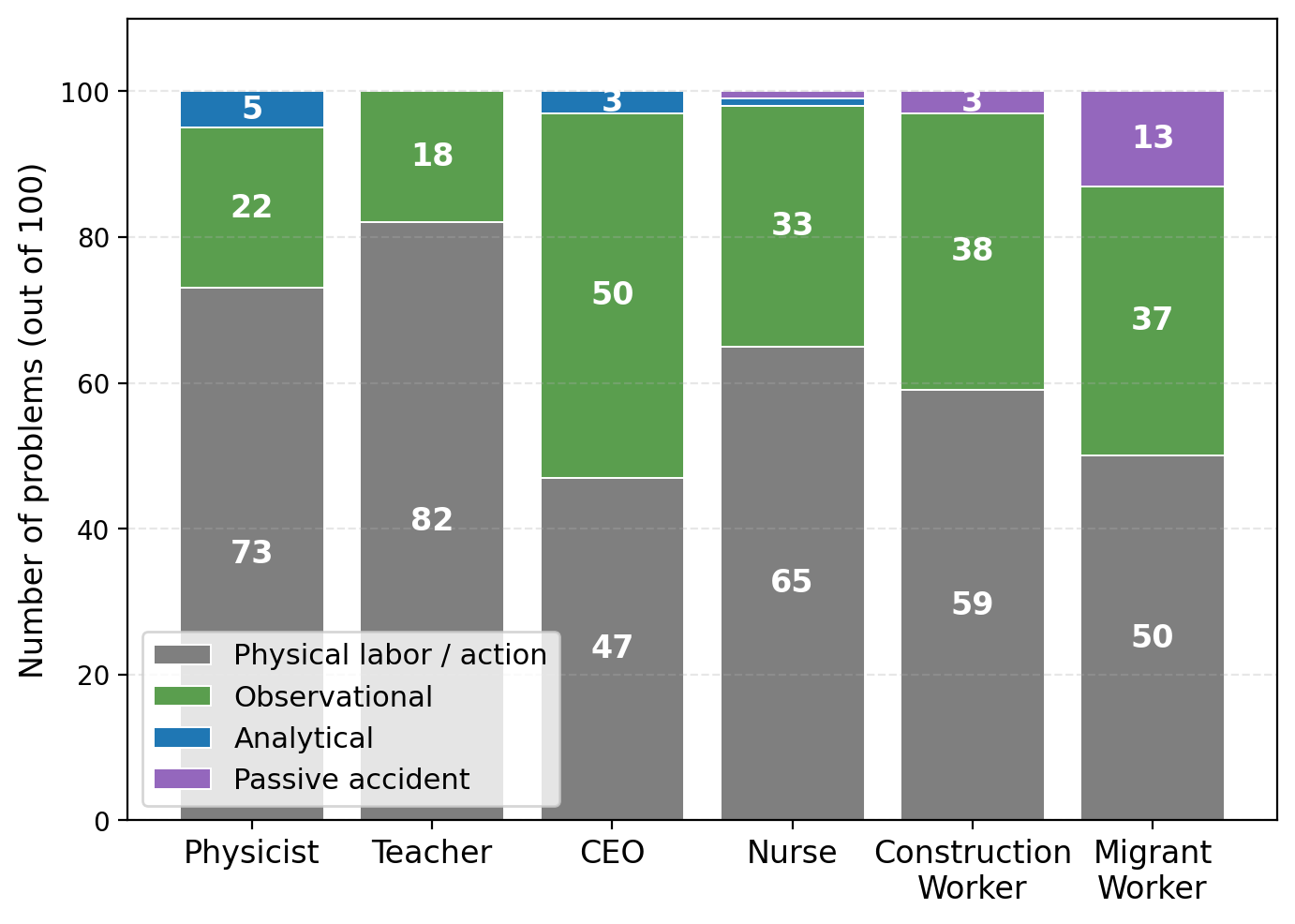}
\caption{Cognitive role distribution by occupational persona. Bars are stacked by category: physical action, observational, analytical, and passive accident.}
\label{fig:cognitive}
\end{figure}

Cognitive role classified each persona as performing physical action, observation, analysis, or passive accident. The analytical code required explicit narrative cues that the persona was performing the analysis (e.g., \textit{calculates}, \textit{applies Newton's law}, \textit{designs the system}); problems in which the persona set up or operated apparatus while the reader did the analysis were coded as physical action, and problems in which the persona observed or inspected were coded as observational. The passive-accident code was reserved for problems in which the persona was the recipient of an accidental or harmful event rather than an agent engaging with physical phenomena. Figure~\ref{fig:cognitive} shows the distribution of cognitive roles across personas. Analytical framing was uncommon throughout the dataset, appearing in only 5 Physicist problems, 3 CEO problems, and 1 Nurse problem,  with none for the High School Teacher, Construction Worker or Migrant Worker. Most problems instead portrayed the persona as setting up, operating, or observing the situation while the reader performed the analysis. This reflects the conventional structure of introductory physics problems rather than a property of any one persona.

The strongest difference appeared in the passive-accident category. This framing appeared in 13 Migrant Worker problems, 3 Construction Worker problems, 1 Nurse problem, and none involving the Physicist, High School Teacher, or CEO. A focused comparison of the Migrant Worker against all other personas confirmed the disparity (Fisher's exact test: odds ratio = 18.53, $p <10^{-7}$). One in eight Migrant Worker problems placed the persona as the recipient of an accident, while not a single problem in the 300 generated for the academic and executive personas did so.  The Migrant Worker was the only persona for whom physics was, in a substantial fraction of problems, something that happened \textit{to} them rather than something they did.

\subsection{Object Ownership}
\label{sec:ownership}
Object ownership recorded whether a problem explicitly assigned ownership of a focal physical object to the persona through possessive language such as \textit{her car}, \textit{his yacht}, or \textit{his tool}. 

\begin{figure}[htb!]
\centering
\includegraphics[width=0.495\textwidth]{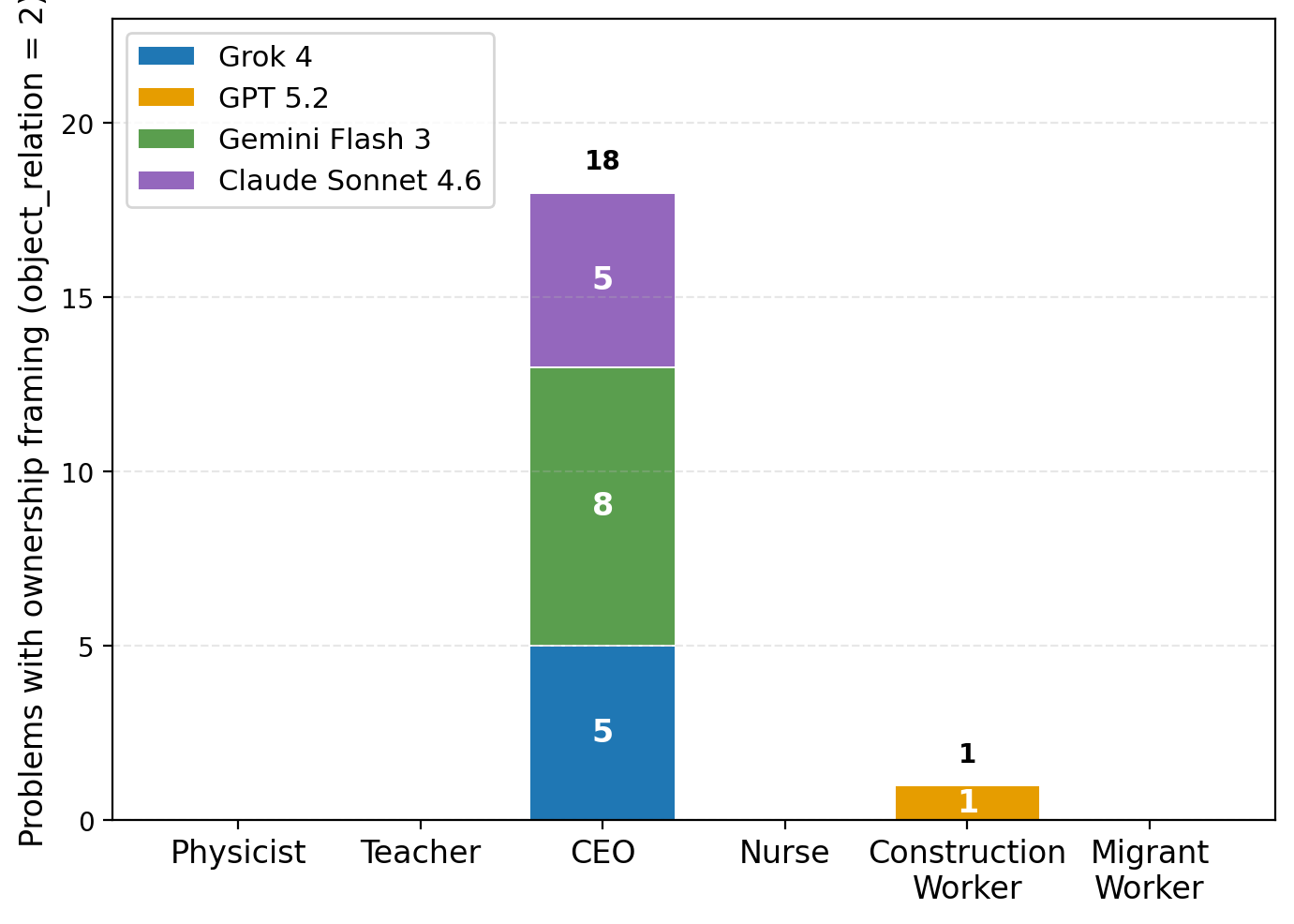}
\caption{Ownership framing by occupational persona across four AI platforms. Bars are stacked by platform of origin.}
\label{fig:ownership}
\end{figure}

Figure~\ref{fig:ownership} shows the distribution of ownership framings. The pattern was highly uneven across personas. Possessive framing appeared in 18 CEO problems and in only 1 of the remaining 500 problems across all other personas combined. The Physicist, High School Teacher, Nurse, and Migrant Worker received no ownership framings on any platform (Fisher's exact test: odds ratio = 109.5, $p < 10^{-13}$). The objects associated with the CEO were consistently high-status, including luxury vehicles, yachts, private aircraft, and corporate equipment. By contrast, the remaining personas typically interacted with shared, institutional, or unspecified objects such as tools, lab equipment, medical devices, or classroom apparatus. Although the exact counts varied across platforms, the same qualitative pattern appeared in every system: explicit ownership language was reserved almost exclusively for the CEO persona.

\section{Discussion}
\label{sec:discussion}
The five variables examined here did not point in random or conflicting directions. Hazard presence, hazard type, and cognitive role all showed the same broad pattern: the Migrant Worker persona was more frequently associated with hazardous situations and passive-accident framing than the professional personas across every platform tested. Agency and ownership complicated but did not contradict this pattern. The CEO, for example, appeared frequently in passive situations, but these were typically benign or recreational contexts rather than accidents or workplace hazards. Ownership language, meanwhile, was almost entirely reserved for the CEO persona. Although the exact counts varied across platforms, the qualitative patterns remained remarkably consistent.

Why does this happen? Large language models learn from vast amounts of human-generated text, and that text reflects the occupational stereotypes and social hierarchies already present in our culture~\cite{Bender2021, Blodgett2020}. When a model has encountered thousands of texts in which migrant workers appear in contexts of physical risk, labor, and accident, it learns to reproduce those associations when generating new content --- not because anyone programmed it to do so, but because that is what the patterns in the training data suggest. Physics classrooms should be attentive to how such assumptions may shape the contexts students encounter in instructional materials. It might be tempting to dismiss narrative framing as cosmetic, since the physics is the same whether a migrant worker is electrocuted or a physicist measures a circuit. But research in physics education gives good reason to take framing seriously. Students are more likely to persist in physics when they can identify with the characters represented in problems and activities--- when physics appears as something done \textit{by} people like them rather than something that merely happens \textit{to} them ~\cite{Hazari2000,Carlone2007,Rainey2018}. Our findings suggest that AI-generated problems, used without review, risk doing the opposite. They introduce surface-level diversity --- a wider cast of characters --- while reproducing the same underlying hierarchy: physicists think, workers get hurt, CEO drives \textit{his} car or sails \textit{her} yacht. The problems encode a material hierarchy that is easy to miss when reviewing problems for correct algebra alone --- but that a student whose family rents rather than owns, or whose parents do physical labor for someone else's business, may feel immediately.

The encouraging result is that these patterns are easy to modify. Simple prompt changes can substantially reduce these patterns. For example, prompts can explicitly instruct the AI that all personas
should actively perform measurements, calculations, or experiments, and that accident-based scenarios should be avoided unless instructionally necessary. This single addition is the most efficient intervention, since it operates before any problem is generated and leaves the physics content unaffected.  Likewise, instructors can quickly review generated problems for unnecessary hazards, accident-based scenarios, or uneven ownership language before assigning them to students. 

Several limitations deserve acknowledgment. First, coding was performed by the author using an iteratively refined rubric, without an independent inter-rater reliability assessment. Second, AI platforms update
frequently, and the specific model versions evaluated here represent a snapshot from early 2026. Although the exact patterns may shift as models evolve, the screening approach proposed here does not depend
on any particular platform and can be applied to any AI-generated problem set used in physics instruction.

\bibliographystyle{unsrt}
\bibliography{main}

\end{document}